\documentclass[aps,prb,groupedaddress, twocolumn]{revtex4}
\usepackage{graphicx} 
\usepackage{dcolumn} 
\usepackage{bm} 
\usepackage{slashed}
\usepackage{subfigure}
\usepackage{natbib}
\usepackage{latexsym,amsfonts,amssymb}
\usepackage{amsmath}
\usepackage[colorlinks, breaklinks=true,linkcolor=cyan, citecolor=cyan,linktocpage=true]{hyperref}
\usepackage[all]{hypcap}

\begin{document}
\title{Effect of site dilution in the two-dimensional attractive Hubbard model}
\author{Saurabh Pradhan}
\author{G. Venketeswara Pai}
\affiliation{Harish-Chandra Research Institute, Chhatnag Road, Jhunsi, Allahabad 211019, India}
\date{\today}
\begin{abstract} 
We study the percolative superconducting transition as the density of randomly placed attractive centers grows in a host metal. Employing the Hubbard-Stratanovich transformation  for  the interaction and  allowing for spatial, thermal fluctuations of the pairing field, we obtain real-space features of the transition from weak to strong coupling. Spectral and  transport properties are studied  in detail. BCS-BEC crossover is discussed in the context of site dilution of attractive centers. 
\end{abstract}
\maketitle

 
\section{Introduction}
The interplay between superconductivity (SC)   and disorder is a long standing problem \cite{belitz,gantmakher,sadowskii,nandini-book}. For weak disorder the two are not expected to be inimical to each other 
since pairing takes place between time-reversed states \cite{anderson} that are present when only potential impurities disorder the system \cite{abrikosov} . In this limit the superconducting state is not 
expected to be much different from the mean field BCS state. In particular, it remains homogeneous 
in low disorder regime. However, when the disorder is strong, it dramatically alters the superconducting phase \cite{amit,dubi}. Large phase fluctuations can reduce the 
superconducting transition temperature from its mean field  value and in a temperature window between $T_c$ and $T_{BCS}$, where $T_c$ is the transition 
temperature and $T_{BCS}$ is the mean field value expected from the BCS theory, a pseudogap phase is expected \cite{sadowskii}. This is a region where 
preformed pairs exist, but global coherence is absent.  Strong disorder also makes the phase highly inhomogeneous. 
Theoretical studies at strong disorder \cite{amit,dubi,mandal,bouadim,feigelman,scalettar,taratDISORDER,trivedi,QMC}  reveal the inhomogeneous natures of the SC state, formation of superconducting puddles and presence of a pseudogap phase where 
long-range SC order  diminishes although the quasiparticle gap remains open. While Josephson effect between puddles can give rise to global SC state\cite{fisher}, strong phase fluctuations
among them may lead to an insulating state\cite{bouadim,tvr}.  Recent experimental studies have probed this behavior \cite{sacepe1,sacepe2,chand,mondal1,mondal2,kamlapure,sacepe3,noat,baturina,kapitulnik}.
They  reveal fragmentation of the superconducting state into islands, pseudogap like features in the normal state, and a change in the normal state 
resistivity suggestive of a metal-insulator transition.\\
~\\
Competition between superconductivity and disorder is  expected to be more interesting in two dimensions since  arbitrarily small disorder is capable of localizing 
electrons \cite{leerama} while the superconducting transition  itself is of Berezenskii-Kosterlitz-Thouless (BKT) type. Experimental studies show a superconductor-insulator transition 
in many two dimensional systems\cite{haviland,hebard,crane,tan}, a theoretical understanding  of which is still not very satisfactory.  Another complication arises as one increases the strength of attractive
 interaction since,  in absence  of disorder, this is expected to lead to a BCS-BEC crossover \cite{nozieresBEC,chen,taratBEC}. The latter results from the Bose condensation of local pairs of electrons arising due to enhanced double  occupancy for large local attractive interactions. The physics is very different from the BCS limit; there is still a large, local pairing gap that is visible in the 
spectral function, but in contrast to the BCS limit, $T_c$ is much reduced and does not scale with the pairing gap, though the zero temperature pairing gap continues to increase with interaction.  Phase fluctuations play a dominant role here  and is the cause of suppression of SC order even when there are strong local pairing tendencies.. The effect of disorder in this limit has not been explored adequately.\\
~\\
Further, there are various systems in which a SC ground state is arrived at by doping an insulating host. For example,  PbTe is a semiconductor, 
but when doped with Tl (Pb$_{(1-x)}$Tl$_x$Te) becomes superconducting beyond a critical $x_c \sim 0.3$ \cite{matsushita1,matsushita2} .  $T_c$ increases with $x$ suggesting that Tl induces pairing.  It is also known that $T_c$ decreases when a superconducting material is doped with certain atoms. Examples include MgB$_2$ doped with carbon \cite{kazakov}
(Mg$_{(1-x)}$C$_x$B$_2$) or aluminium (Mg$_{(1-x)}$Al$_x$B$_2$) \cite{karpinski}. A simple way of looking at this problem is to assume that a host system is 
doped with inhomogeneous attractive centers which promote local pairing\cite{note1}. As the number of such attractive centers increases, superconducting islands start to form. 
However, onset of superconductivity requires percolation of  these puddles, thereby establishing global phase coherence. This problem has several interesting features. 
Disorder and superconductivity contribute on an equal footing and the superconducting state is expected to be  intrinsically inhomogeneous. In absence of attractive centers, 
the host could be metallic or nonconducting, though we study only the former in this paper.
One could also explore the BCS-BEC crossover in the context of dilution of attractive centers  if one is able to handle the regime of large attractive interactions.\\ 
~\\
Theoretical  studies based on the above picture have been carried out previously using mean field theory \cite{litak,Meir,aryanpour1,aryanpour2} or quantum Monte Carlo (QMC)\cite{scalleter}. The former does not have the prospect of studying 
large interaction strengths. The latter can handle the entire range, but is numerically expensive with obvious system size limitations; 
transport is harder to evaluate.  Recently, dynamical mean field theory (DMFT) \cite{georges} was employed in conjunction with coherent potential approximation (CPA)  to treat disorder\cite{shenoy,raja}. 
However, neglect of spatial correlations lead to unphysical results at strong interactions.
To this end, we use a numerically less expensive method that captures the entire parameter regime  while retaining  thermal fluctuations of the pairing field
and is able to shed light on the physics in real space.  
We employ the random attractive Hubbard model \cite{ranninger} where the number of attractive centers is determined by the dilution on an average. 
We use a real-space Hubbard-Stratanovich (HS) \cite{dubi,Meir,erez,dupuis}
transformation by introducing auxiliary fields in the pairing and charge channels that couple to electrons. For simplicity, we assume the auxiliary  fields to be  classical; 
while we allow spatial fluctuations of amplitude and phase of the pairing field, we neglect their dynamics.  This results in studying the self-consistent quantum dynamics of electrons coupled to  thermally fluctuating, classical pairng field which is treated numerically using Monte Carlo method\cite{mayr} . The details of the model and numerical procedure are  given in Section II.  
We discuss the critical temperature of the superconducting transition and its variation with dilution and strength of interaction in
Section III.  Afterwards, we present spectral and transport properties of this model. Being a real-space method, this gives us a direct image of the physics in real space, 
while allowing to access the BCS-BEC crossover regime. We conclude by pointing out certain limitations of the present approach and possible extensions.

\section{Model and The Static Auxiliary Field Method}
To study the nature of percolative  superconductivity due to variation in the density of attractive centers, we employ a minimal model, which is the attractive Hubbard model with site dilution,  that captures the essential features of the problem. The Hamiltonian employed is 
\begin{equation}
H = -t \sum_{\left< ij \right>, \sigma} c^{\dag}_{i \sigma}  c_{j \sigma} - \sum_i  U_i n_{i \uparrow} n_{i \downarrow} - \mu \sum_i n_i.
\end{equation}
Here, $t$ is the nearest-neighbor hopping integral (which we take to be unity to set the energy scales), $U_i$ is the  strength of  attractive interaction that
 is site-dependent, $\mu$ is the chemical potential which fixes the mean electron density.  In this paper, we fix the electron density to be $n = 0.875$. 
 However, the physics is not very sensitive to changes in average electron density, except at half filling.
The case of half filling is special, which we discuss in the last section. Site dilution is introduced {\it via} site dependent $U_i$  
that follows a bimodal probability distribution  such that $U_i = U$ with probability $P(U)  = \delta $ and $U_i = 0$ with probability 
$P(U) = 1-\delta$ \cite{note2}, where $\delta$ is the average number of sites having attractive centers (e.g., $\delta =1$ 
when all sites have attractive centers, which is the clean limit). We employ the Hubbard-Stratanovich transformation to reduce the interacting, quartic Hamiltonian
to a quadratic fermionic Hamiltonian  coupled to a pairing field $\Delta_i$, which is a complex variable and a real, scalar-valued charge (or, equivalently density)  field $\phi_i$. 
The resulting Hamiltonian reads : 
\begin{eqnarray}
H_{eff} = &-& t \sum_{\left< ij \right>,\sigma} c^{\dag}_{i \sigma} c_{j \sigma} - \mu \sum_i n_i + \sum_i \left( \Delta_i c^{\dag}_{i \uparrow} c^{\dag}_{i \downarrow} + h.c. \right)
\nonumber \\
& +& \sum_i  {{{\left| \Delta_i  \right| }^2} \over U_i} + \sum_i \phi_i n_i + \sum_i {\phi_i^2 \over U_i},
\end{eqnarray}
where $\Delta_i = \left< c_{i \uparrow} c_{i \downarrow} \right>$ and $n_i = \sum_{\sigma} c^\dag_{i \sigma} c_{i \sigma}$. The partition function can be evaluated in terms of the effective Hamiltonian and is given by
\begin{equation} 
{\cal {Z}} = \int {\cal{D}} \Delta {\cal {D} }\Delta^* {\cal {D}} \phi {\cal {D}} \left[ c^\dag,  c \right] e^{-\beta H_{eff}},
\end{equation}
so that the probability of occurrence of a particular configuration of $\Delta_i$ at inverse temperature $\beta = 1/\left( k_B T \right)$ is obtained from
\begin{equation}
P \left( \Delta_i \right) = {1 \over {\cal{Z}}}  \int   {\cal {D}} \phi {\cal {D}} \left[ c^\dag,  c \right] e^{-\beta H_{eff}}.
\end{equation}

The saddle-point solutions of the action corresponding to the effective Hamiltonian give Bogoliubov-de Gennes (BdG) equations for the pairing field $\Delta_i$ and the charge field $\phi_i$. 
While at this level the action is exact,  to make progress, we assume that the pairing fields are static (i.e., we neglect quantum fluctuations), 
but their amplitudes and phases are site-dependent and thermally fluctuating \cite{taratBEC,taratDISORDER,mayr} . Charge field is also assumed to be classical. At finite temperatures, 
this necessitates thermally averaging over their most probable configurations, which we carry out using a Monte Carlo (MC) estimation of their weights based on Metropolis algorithm.
This essentially means that for a given electron density and temperature, we start with a random configuration of attractive centers  by fixing the amount of site dilution, 
and a judicious choice of the pairing and charge fields at every site. This leaves us with a problem of  electrons moving in random (classical) fields that requires 
an exact diagonalization of the fermion problem. A thermal sampling of the most probable configurations of the auxiliary fields is performed by Monte Carlo updating of the (classical)  fields.
Thermodynamic properties of the system as well as spectral features of electrons and transport  are obtained by averaging over configurations thus obtained.
This method, which requires exact diagonalization of the electron system at every Monte Carlo step, obviously restricts the system size and to circumvent it 
we use  a traveling cluster algorithm (TCA)\cite{TCA} . Here, the fermion problem is diagonalized on a smaller cluster around the chosen MC update site, 
embedded in a much larger lattice. The cluster moves during every MC update restoring ergodicity. A similar approach was recently used successfully for the case of repulsive Hubbard-Holstein model in two dimensions \cite{holstein}.

~\\

Before presenting our results, we review the previous works based on the above model.  These include mean field calculations based on BdG equations\cite{litak,aryanpour1,aryanpour2} with disorder treated 
using CPA, quantum Monte Carlo\cite{scalleter} , and the dynamical mean field theory with iterated perturbation theory (IPT)  
as an impurity solver in conjunction with CPA to handle disorder\cite{raja,shenoy}. In general, a critical concentration of attractive centers, $\delta_c$, is required to get superconducting ground state. 
The system undergoes a first order metal-SC transition at $\delta_c$. While $\delta_c$ increases with $U$ in mean field calculations, it is seen to decrease and then saturate with $U$  in QMC.
DMFT studies reveal that $\delta_c$ decreases sharply with increasing $U$. 
For all $U \geq$  2.7, $\delta_c \sim$ 0. This is obviously an artifact of the infinite coordination number employed in DMFT, neglecting spatial correlations. They also find that suppressing dynamic fluctuations leads to $\delta_c =$ 0, suggesting that arbitrary small number of sttractive dopants are needed for onset of superconductivity.  
In general, $\delta_c$ displays a strong dependence on $U$, which suggests that the transition from the metallic to SC state  cannot be thought of entirely in terms of percolation alone.
In the next section, we provide results on the metal-SC transition in the model based on our study focusing on the variation of structure factor with temperature which determines
$T_c$.

\section{Order Parameter and Critical Temperature}
Once the system reaches equilibrium, we use the thermal averaged structure factor for the pairing field  $\Delta_i = \left<  c_{i \uparrow} c_{i \downarrow} \right>$,  to track the onset of superconductivity  as a function of dilution and temperature: 

\begin{equation}
S\left( {\bf q} \right) = {1 \over N^2} \sum_{ij}  \left< \Delta_i \Delta^*_j  \right> e^{i {\bf q} \cdot \left( {\bf r}_i - {\bf r}_j \right)}.
\end{equation}

For a uniform superconducting solution, we look at the wave vector ${\bf q} = \left( 0, 0 \right)$ and the corresponding structure factor $S( {\bf 0})$.  For a given dilution, $S( {\bf 0})$ is vanishingly small at high temperatures 
and  starts picking up at a characteristic temperature, which we identify with the superconducting transition temperature $T_c$.  
A typical result is plotted in Fig. 1 for $U=8$.   We note that there is a critical concentration of attractive centers, $\delta_c$, 
needed to have "global" superconductivity, which for this case happens to be roughly $\delta_c = 0.6$.   Further, the saturation value and $T_c$ increases with $\delta$. 
As we will discuss later, the onset of superconductivity is brought about by  percolation of locally superconducting islands and having larger number of attractive centers helps in enhancing
 the superconducting correlations, and hence the transition temperature itself.  For most of our discussions, we have used a system size of 32 $\times$ 32 
 with the size of the traveling cluster being 8 $\times$ 8.  There is a marginal decrease in $T_c$ as system size increases which is to be expected  in a two-dimensional system. 
 However, we expect that in a three dimensional system, even with a small hopping between layers (or, in other words a large anisotropy between them), 
 transition temperatures would stabilize. Of course, we do not take up this task since it is computationally expensive and more importantly, we are interested 
 in the generic features of the problem, demonstrating the usefulness of the procedure. Results for other values of $U$ show similar behavior. 
 However, a notable change is the non-monotonic variation of $T_c$ as a function of $U$ which we discuss next.\\
 ~\\
 \begin{figure}[htp]
\includegraphics[scale=0.45]{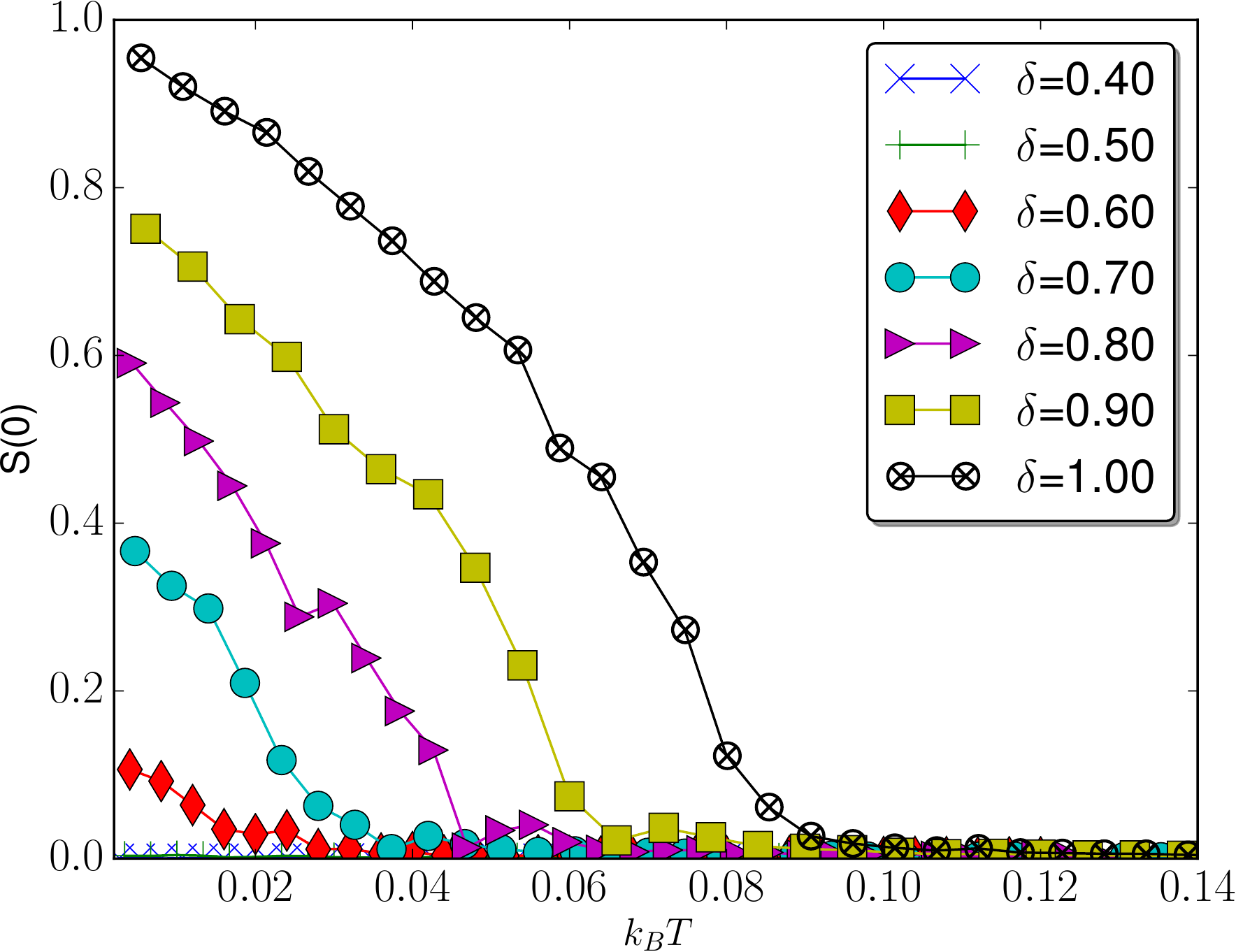}
\caption{ (Color online) Structure factor $S({\bf  0})$ for the pairing field as a function of temperature for $ U =$ 8 and for different values of dilution $\delta$.}
\end{figure}

\begin{figure}[htp]
\includegraphics[scale=0.45]{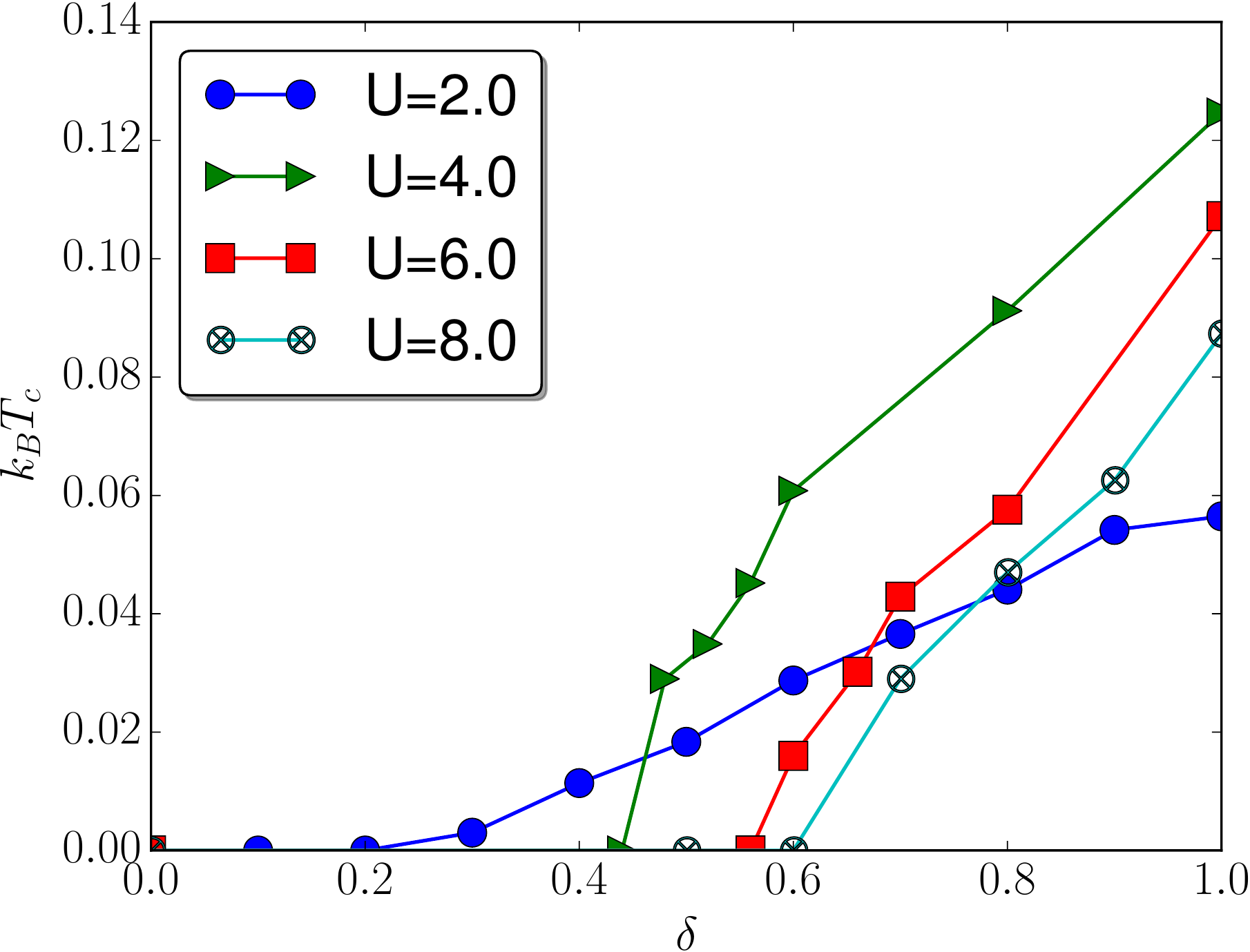}
\caption{ (Color online) Transition temperature $T_c$ as a function of dilution $\delta$ for different values of interaction strength $U$.}
\end{figure}

Fig. 2 shows the variation of $T_c$ as a function of $\delta$ for different values of $U$.  The critical density of attractive centers needed for onset of 
"globally phase coherent" superconductivity increases  with $U$ and appears to saturate around $U \sim 6$.  For small values of $U$,  $T_c$ is determined 
by the pairing scale at which the amplitude of the Cooper pair become non-zero. This is the BCS limit where phase fluctuations hardly play any role.  
However, as $U$ increases, the pair size (or equivalently, the coherence length) comes down and onset of superconductivity is determined 
by the phase coherence temperature, instead of the pairing scale.  In fact, at large $U$,  the pairing amplitude remains almost constant across the transition 
at the sites where there is an attractive center. However, the relative phases among the sites fluctuate wildly and global coherence is established
only at a very low temperature (compared to the BCS mean field value) and is determined by  phase fluctuation scale which goes as $\sim t^2/U$.  This results in a nonmonotonic variation of $T_c$ with $U$ arising due to BCS-BEC crossover and  is most clearly seen in Fig. 2 for $\delta=$ 1 (the clean limit). However, such a behavior sets in
roughly at $U =$ 6, establishing this crossover in an intrinsically disordered system. This also signals a clear separation of energy scales. The zero-temperature pairing gap in the quasiparticle spectrum continues to increase with $U$, though it determines the superconducting $T_c$ only in the weak-coupling limit. This also results in a nontrivial behavior of the normal phase, wherein it changes from a Fermi liquid to a gapped phase at large interaction strengths.
There is a smooth crossover 
between these two regimes with an intermediate high-temperature normal phase that intervenes in the crossover region with anomalous properties, 
which we will discuss in the next section. 
~\\

Our method  incorporates spatial fluctuations of the pairing field, both its amplitude and phase, in an unbiased way and in fact, this is a crucial ingredient  
to obtain the BCS-BEC crossover. The latter arises due to site-dependent phase fluctuations, which cannot be captured in the 
conventional BCS framework as was discussed earlier. Further, unique to this problem s the local charge fluctuations that can be quite large due to site dilution.  
Next, we discuss the role of each of these, the charge and the amplitude  and phase of pairing field fluctuations and their role in nucleating/stabilizing  superconductivity 
as a fundtion of temperature and dilution. In the clean limit with $\delta = 1$, all sites have uniform charge distribution and  fluctuations are negligible. 
However, as sites are being diluted, there is a strong tendency to have average charge density to be larger near attractive centers and this can be seen most clearly in the lowest rows 
of Fig. 4 corresponding to $U =$ 8.\\
~\\

\begin{figure}[htp]
\includegraphics[scale=0.45]{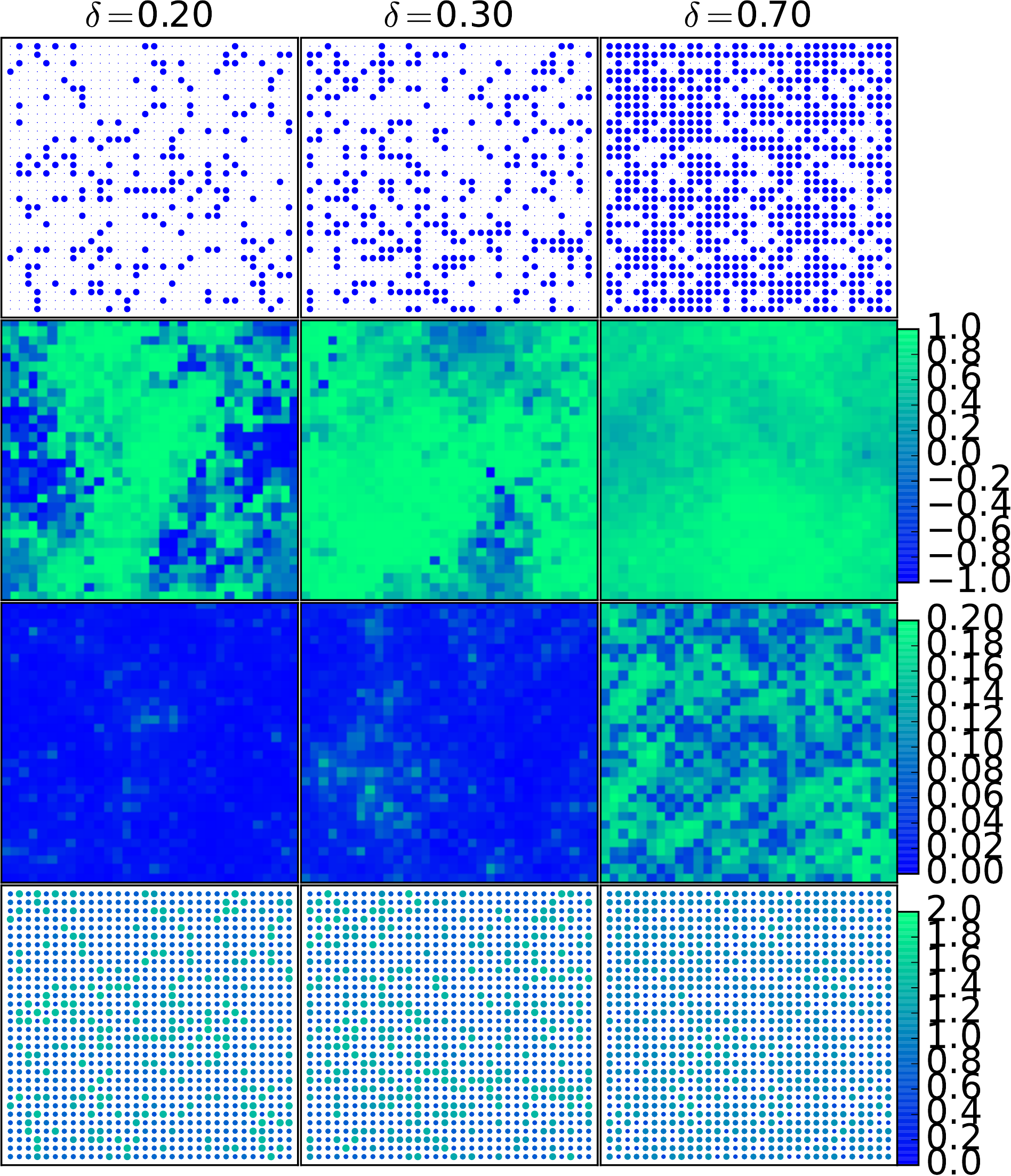}
\caption{(Color online) Real-space configurations of the distribution of attractive centers and various auxiliary fields  at $U =$ 2 at the lowest temperature ($T=$ 0.001 in units of $t$.
The three columns correspond to different dilution : $\delta =$ 0.2 (first), $\delta =$ 0.3 (second) and $\delta =$ 0.7 (third). The first row gives the distribution of attractive centers with blue circles denoting sites with $U_i = U$. The other rows depict the phase $\cos (\arg (\Delta_i))$ (second) and the amplitude $\left| \Delta_i \right|$ (third) of the pairing fields, and the charge field $n_i$ (fourth). The system size is 32 $\times$ 32.}
\end{figure}

On the contrary, the local amplitude of the pairing field $\left| \Delta_i \right|$, where $\Delta_i  = \left| \Delta_i \right| e^{i \theta_i}$,  is almost vanishing at every site 
for large enough dilution, except in small islands where it is nonzero, grows in size  as dilution decreases. This in fact, is the origin of the percolative nature of the transition 
as a function of  $\delta$ for a fixed $U$. There are puddles where amplitude is nonzero, but there are large intervening regions where it is vanishingly small. 
There is phase coherence within a given puddle, but that cannot stabilise a global superconducting state.  Beyond a percolation threshold $\delta_c$,  
which happens mostly in the BCS-BEC crossover region, there is sufficient  pairing amplitude at most sites since the dilution is less.  
However, physics in this strong-coupling region is determined entirely by phase fluctuations. To bring out this feature more clearly, we present the same physical quantities in Fig. 4 for
$U =$ 8.  The behaviour of $\left< n_i \right>$ and $\left| \Delta_i \right|$ are qualitatively different at larger values of $U$.  
The site-to-site charge fluctuations increase enormously, with attractive centers having localised pair of electrons with opposite spins. 
As $\delta$ varies from 0.4 to 0.8, charge-rich sites increases in number. The charge density at the charge-rich sites reaches close to 2 for systems with large $U$.
Amplitude of the order takes  zero  almost everywhere in the lattice for $\delta=0.40$ or smaller than that.  It starts to take non-zero values from $\delta=0.50$ onwards. 
However, there is a strong fluctuation of these amplitudes compared to $U=2$. Average value of this amplitude on sites with $U_i=U$ increases  with $\delta$ for $\delta>0.50$ 
whereas $|\mathbf{\Delta}|$ does not change much on the sites with   $U_i=0$.\\ 
~\\

Naturally, the picture that emerges is that, as expected, fluctuations of the phase degrees of freedom  do not play a major role for small values of $U$. The transition is  entirely BCS-like. Variation of $\delta$ affects the percolative nature of the transition since, the otherwise locally phase coherent islands have to overlap to  give rise to a globally superconducting state.  However, as can be seen from the second rows of Figs. 3 and 4, the phase of the order parameter changes dramatically 
as we change $U$. At large $U$, even though the amplitudes have acquired reasonably large values at every site, their phases become uncorrelated due to 
thermal fluctuations of these soft degrees of freedom. This reduces $T_c$ as $U$ increases.  The nonmonotonic variation of $\delta_c$ and $T_c$ as a function of $U$ suggests that
the transition from a metallic to a superconducting ground state cannot be thought of as entirely due to percolation of amplitude puddles, as noted in an earlier work\cite{scalleter} .  The contrast between the two extreme ends of weak and strong coupling is striking:  There are phase-coherent patches of relatively small amplitudes
 at small $U$ and leads to a percolative transition as the number of attractive centers increases; however,  strong phase fluctuations suppress the $T_c$
at large $U$ even though the pairing amplitude is quite strong enough at most of the sites.\\
~\\

\begin{figure}[htp]
\includegraphics[scale=0.45]{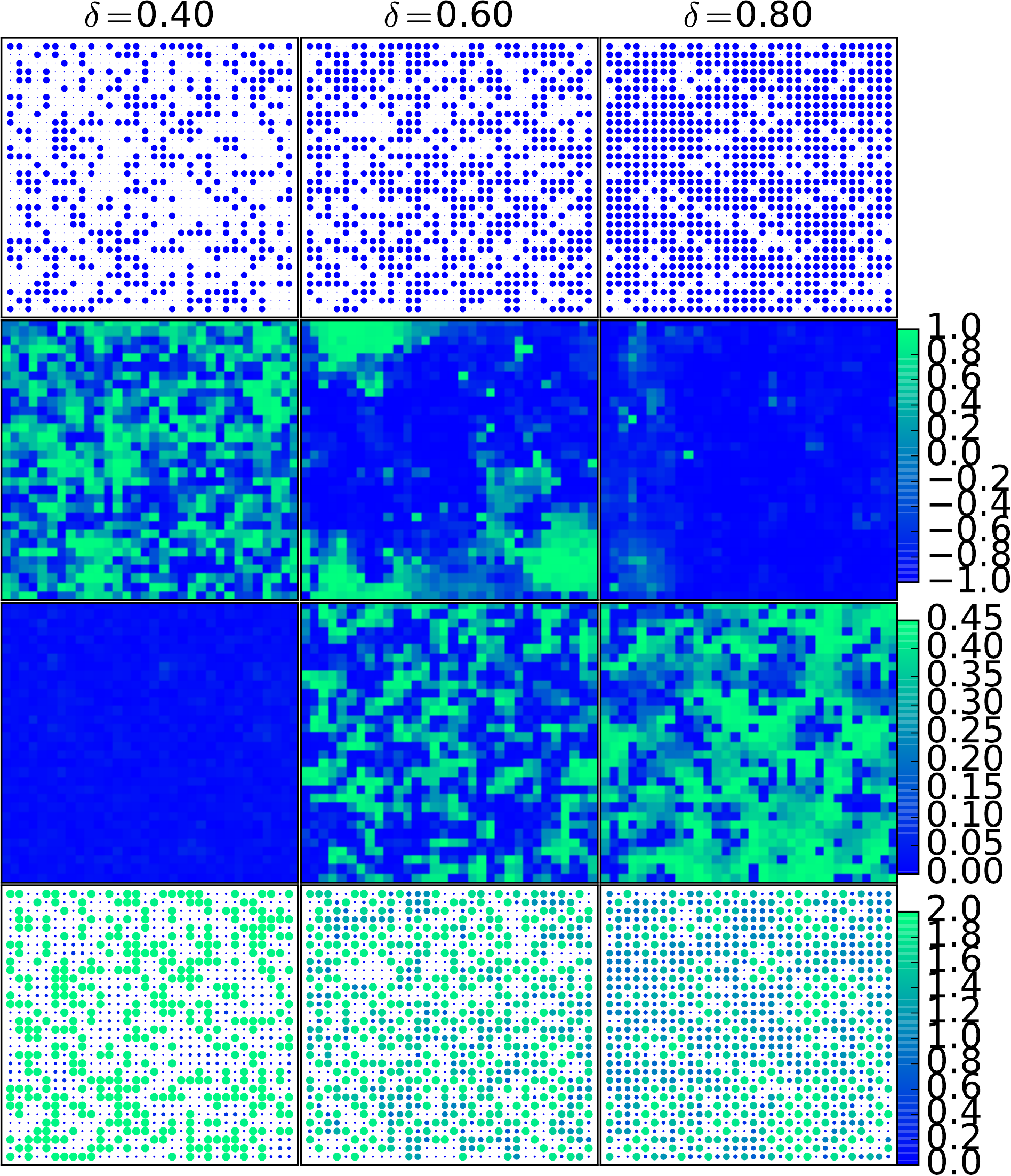}
\caption{(Color online) Real-space configurations of the distribution of attractive centers and various auxiliary fields  at $U =$ 8 at the lowest temperature ($T=$ 0.001 in units of $t$.
The three columns correspond to different dilution : $\delta =$ 0.4 (first), $\delta =$ 0.6 (second) and $\delta =$ 0.8 (third). The first row gives the distribution of attractive centers with blue circles denoting sites with $U_i = U$. The other rows depict the phase $\cos (\arg (\Delta_i))$ (second) and the amplitude $\left| \Delta_i \right|$ (third) of the pairing fields, and the charge field $n_i$ (fourth). The system size is 32 $\times$ 32.}
\end{figure}

\section{Electronic Spectral Functions}

\begin{figure}[htp]
\includegraphics[scale=0.45]{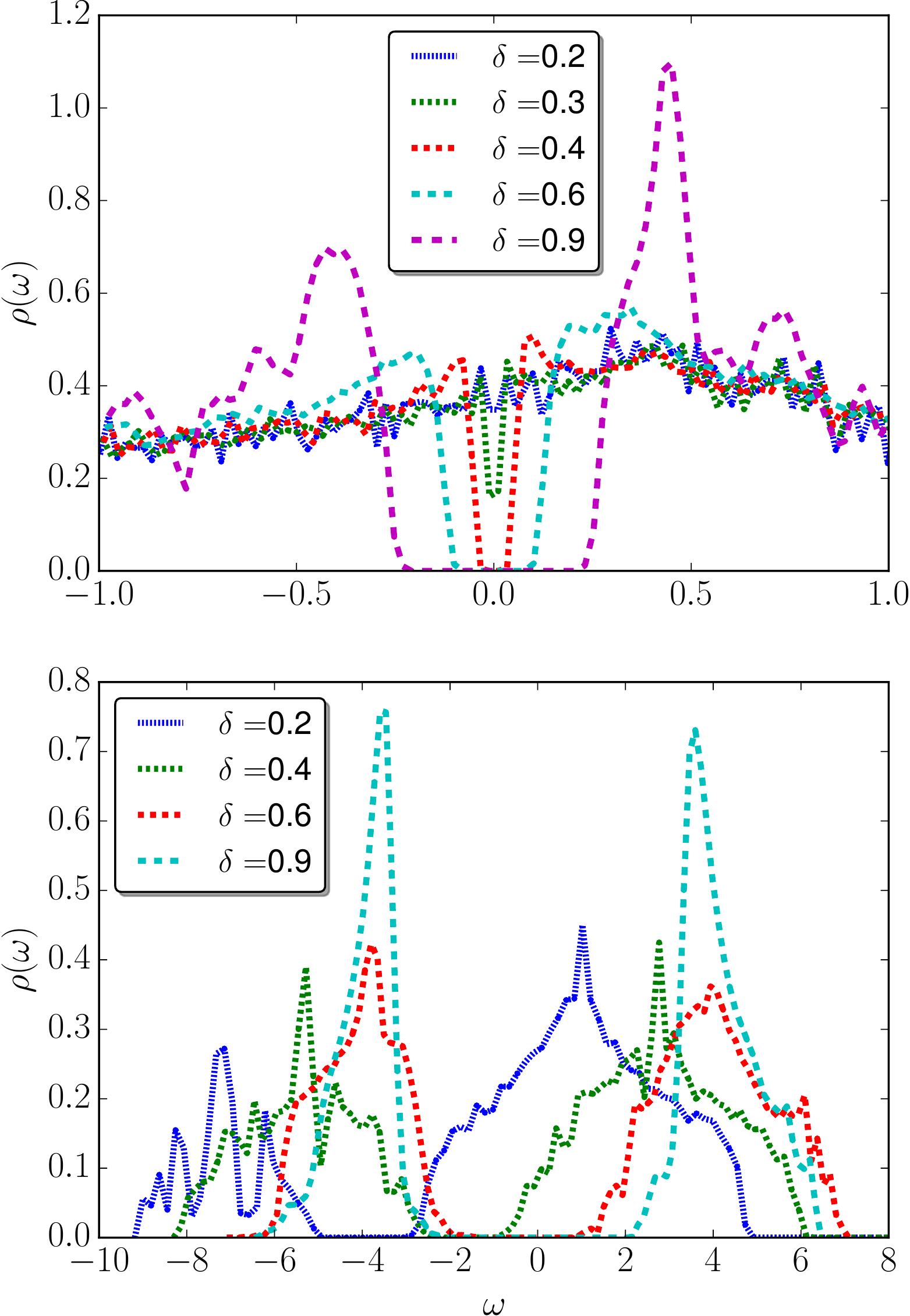}
\caption{(Color online) Single-particle density of states at the lowest temperature ($T =$ 0.001 in units of $t$) as a function of dilution for (a) $U =$ 2 and (b) $U =$ 8.}
\end{figure}

In this section we look at the spectral properties of the system, in particular, concentrating on the single particle density of states. This is expected to show a bulk 
superconducting gap and coherence peaks when the system turns superconducting. Fig. 5a shows its variation as a function of $\delta$ at the lowest temperature 
we have accessed in the weak-coupling regime with $U =$ 2.. There is no gap until about $\delta \sim$ 0.3 and a gap appears as $\delta$ approaches 0.4. A clear spectral gap is seen at 
larger values of $\delta$, which increases with increase in $\delta$. Also visible are the coherence peaks on either side of the gap. These results match very well with those obtained from the structure factor in Fig. 1.
 In fact, this gap vanishes as we increase the temperature across $T_c$ in this region of parameter space (for small $U$).  However, the temperature variation of the spectral gap changes dramatically as we increase $U$. The temperature at which the gap vanishes increases very rapidly with $U$ even though as mentioned in the previous section, the $T_c$ comes down drastically. This shows a clear separation the pairing scale determined from the spectral gap  and superconducting scale that determines $T_c$.

~\\

We also find that the gap in the density of states is larger at larger values of $U$, as expected. However, the effect of $\delta$ is more subtle.  
While the gap decreases with $\delta$, it exists even when superconductivity is not present in the system. We show typical results for a representative value of 
$U = 8$ in Fig. 5b.  Even when very few sites have attractive centers, when $U$ is large enough,   local pairing tendencies are stronger. 
Thus centers which have large $U$ become doubly occupied and gain an energy of the order of $-U$. 
In the large $U$ limit this states with energy $-U$ will be isolated from the 
kinetic energy band. If the average no of particle is nearly half then one has to fill up higher 
energy states above the gap. Density at state at the Fermi energy($\omega=0$) will be non-zero; the ground state of the system would be a metal as can be seen for $\delta =$ 0.2 and 0.4.
Fig. 6 shows the behavior of single particle density of states for two different dilution $\delta =$ 0.5 (Fig.  6a) and 0.8 (Fig. 6b) at two representative temperatures. As expected, 
sharp coherence peaks are not seen in the spectral function. What is striking is that at larger $\delta$, the spectral gap, even though diminished in size, persists at high temperatures

\begin{figure}[htp]
\includegraphics[scale=0.40]{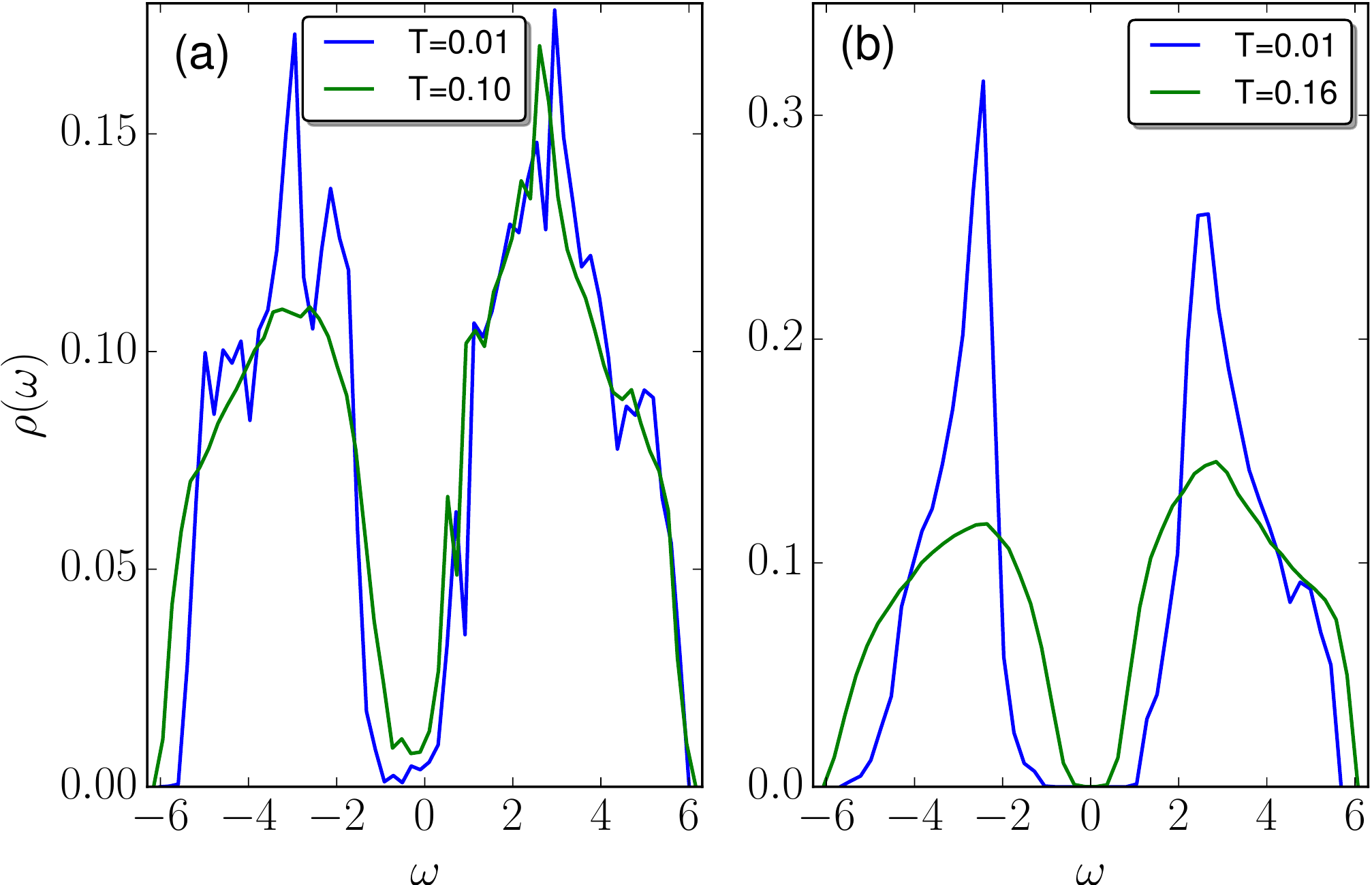}
\caption{(Color online) Single-particle density of states for $U =$ 6 with $\delta =$ 0.5 (a)  and 0.8 (b) at different temperatures.}
\end{figure}

\section{Optical Transport}

\begin{figure}[htp]
\includegraphics[scale=0.45]{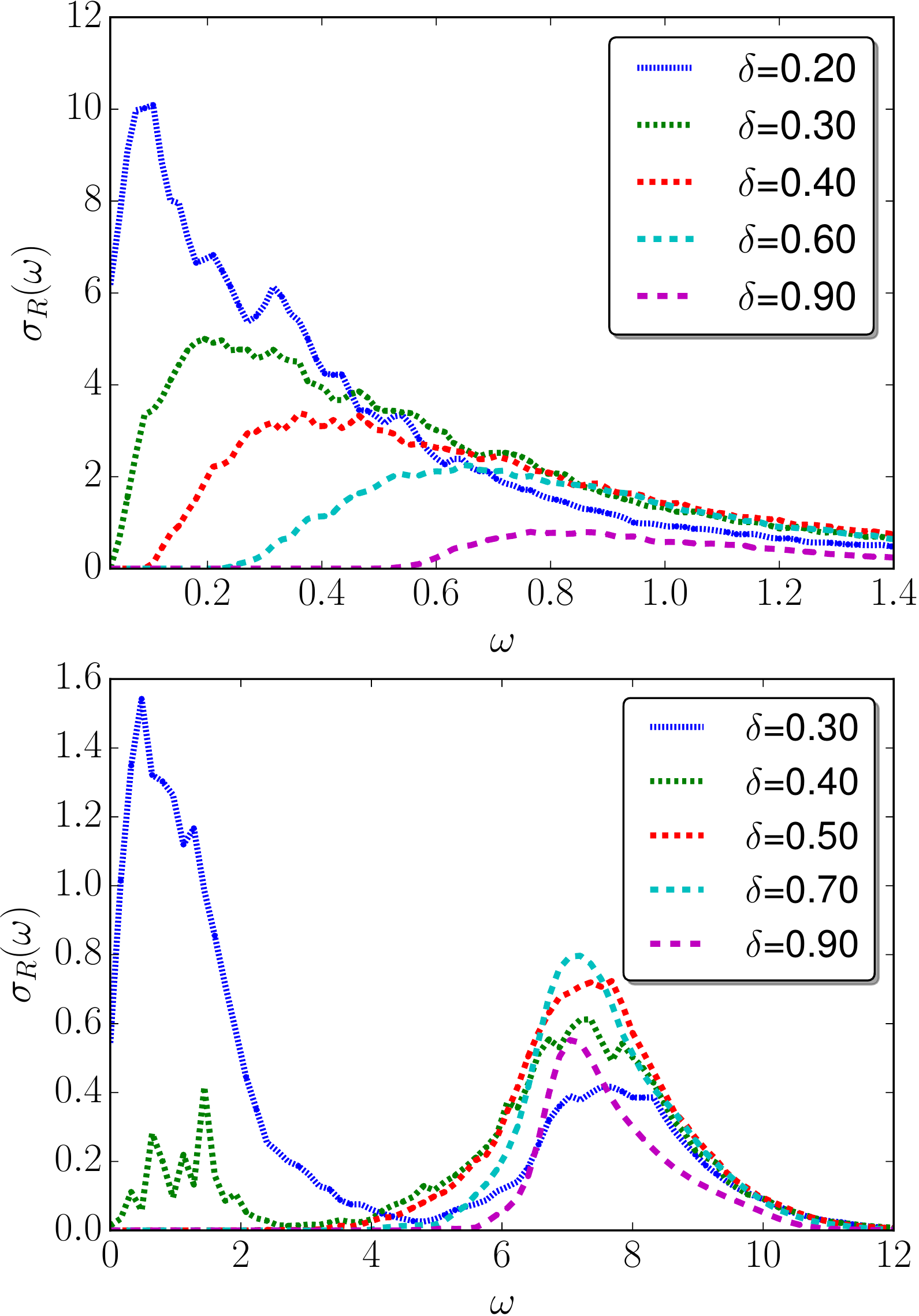}
\caption{(Color online) The real part of the optical conductivity $\sigma_R(\omega)$ as a function of frequency $\omega$ at the lowest temperature ($T =$ 0.001 in units of $t$) for different values of site dilution and interaction strengths $U =$ 2 (upper panel) and $U =$ 8 (lower panel). }
\end{figure}

Optical conductivity is expected to be directly correlated to the spectral features discussed previously and we explore it in this section.
In the superconducting state, $\sigma(\omega)$ has two contributions; there is a zero frequency diamagnetic response that is proportional to the superfluid stiffness.
 In addition, there is a  $\omega$-dependent regular part arising due to various effects such as pair breaking, quasiparticle scattering etc.
 We concentrate on the latter in this section.
 In the superconducting ground state, in the BCS limit, the latter contributes only at frequencies larger than twice the superconducting gap. At nonzero temperatures,
  there is finite contribution even inside this frequency window, but exponentially suppressed. \\
~\\
For small values of $U$, the above features appear to be generic and a representative behavior is shown in Fig. 7a for $U =$2. The gapped spectrum results in the vanishing of the optical conductivity for $\delta \geq$ 0.4. At small $\delta$, the system remains a metal. However, there is intrinsic disorder present due to small number of attractive centers which give rise to enhanced scattering even at low temperatures and results in a non-Drude behavior of $\sigma(\omega)$.
In the large $U$ limit optical conductivity behaves differently as a function $\delta$. Strong optical response at larger $\omega$ arises due to excitations across the SC gap, while the excitations within the kinetic band, separated from the lower band at $-U$, give rise to low frequency non-Drude-like behavior. (See Fig. 7b.)\\
~\\

\begin{figure}[htp]
\includegraphics[scale=0.45]{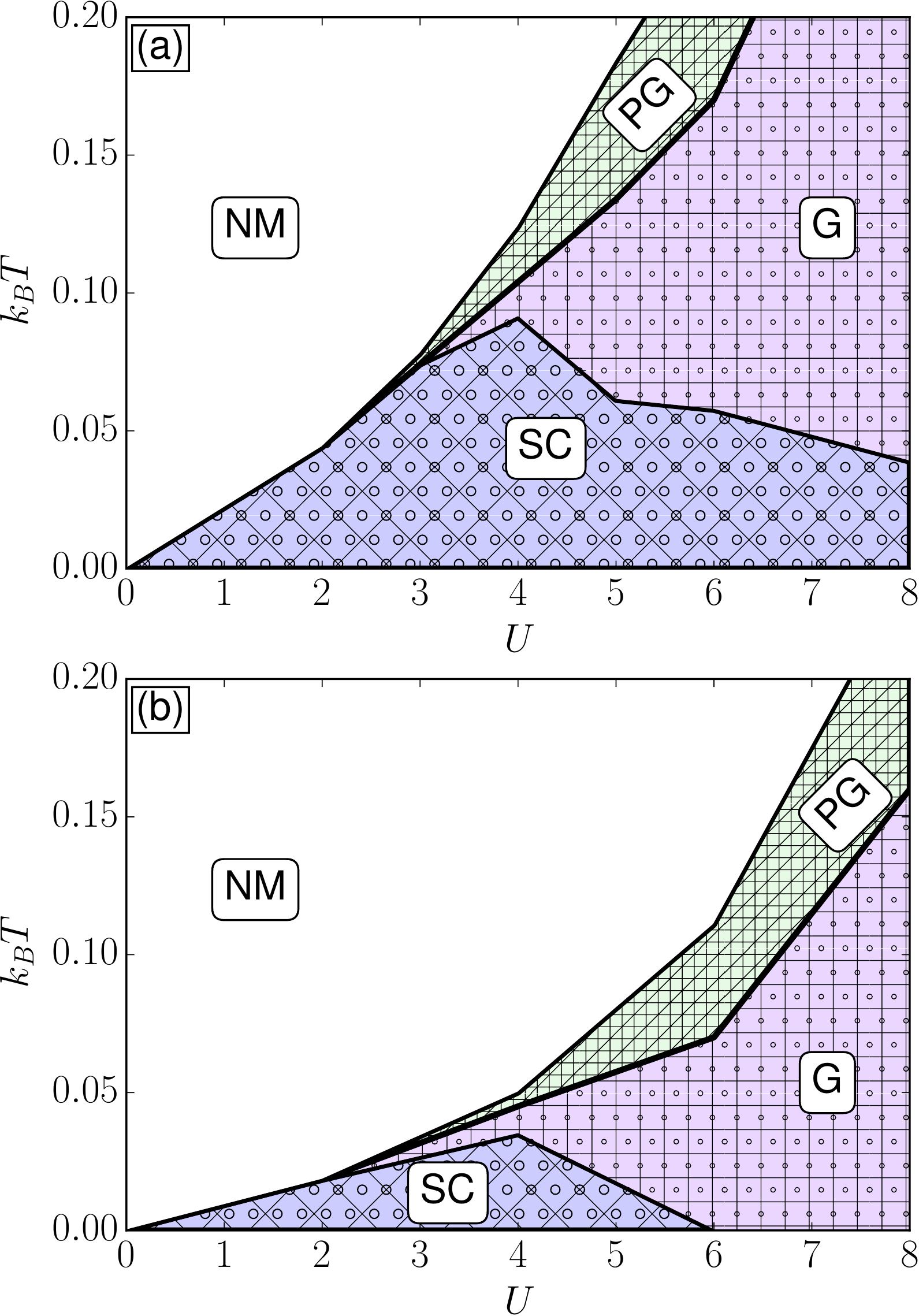}
\caption{(Color online) Two representative phase diagrams of the site-diluted attractive Hubbard model as a function of temperature $T$ and strength of the attractive interaction $U$ for (a) $\delta =$ 0.8
and for (b) $\delta =$ 0.5. SC, NM, PG, and G represent superconducting, normal metal, pseudogap phase, and gapped phases.} 
\end{figure}

We summarise these findings in the next two figures. In Fig. 8, we give two representative phase diagrams of the site-diluted attractive Hubbard model  as a function of
$T$ and $U$ for two specific values of $\delta =$ 0.8 (Fig. 8a) and $\delta =$ 0.5 (Fig. 8b). At small values of $U$, the system turns from a BCS-like superconducting state to a normal metal above $T_c$, which scales with the pairing gap.  However, at larger $U$ even though the pairing gap continues to increase, $T_c$ reduces from the mean field value due to strong phase fluctuations. The normal state above $T_c$ has quasiparticle gap in the spectrum. However, there is an intervening region, where,  a hard gap does not appear in the spectrum, but there is significant reduction of spectral weight at low frequencies. We call this the pseudogap phase.
For smaller $\delta$, SC ground state vanishes above a critical value of $U$ since there is a critical dilution $\delta_c$ that increases with $U$. Finally, Fig. 9 shows the three dimensional phase diagram of the model as a function of $T$, $\delta$, and $U$, clearly brining out the variantion of $\delta_c$ with $U$ and the nonmonotonic behavior of $T_c$ as the strength of interaction changes. \\
~\\

\begin{figure}[htp]
\includegraphics[scale=0.55]{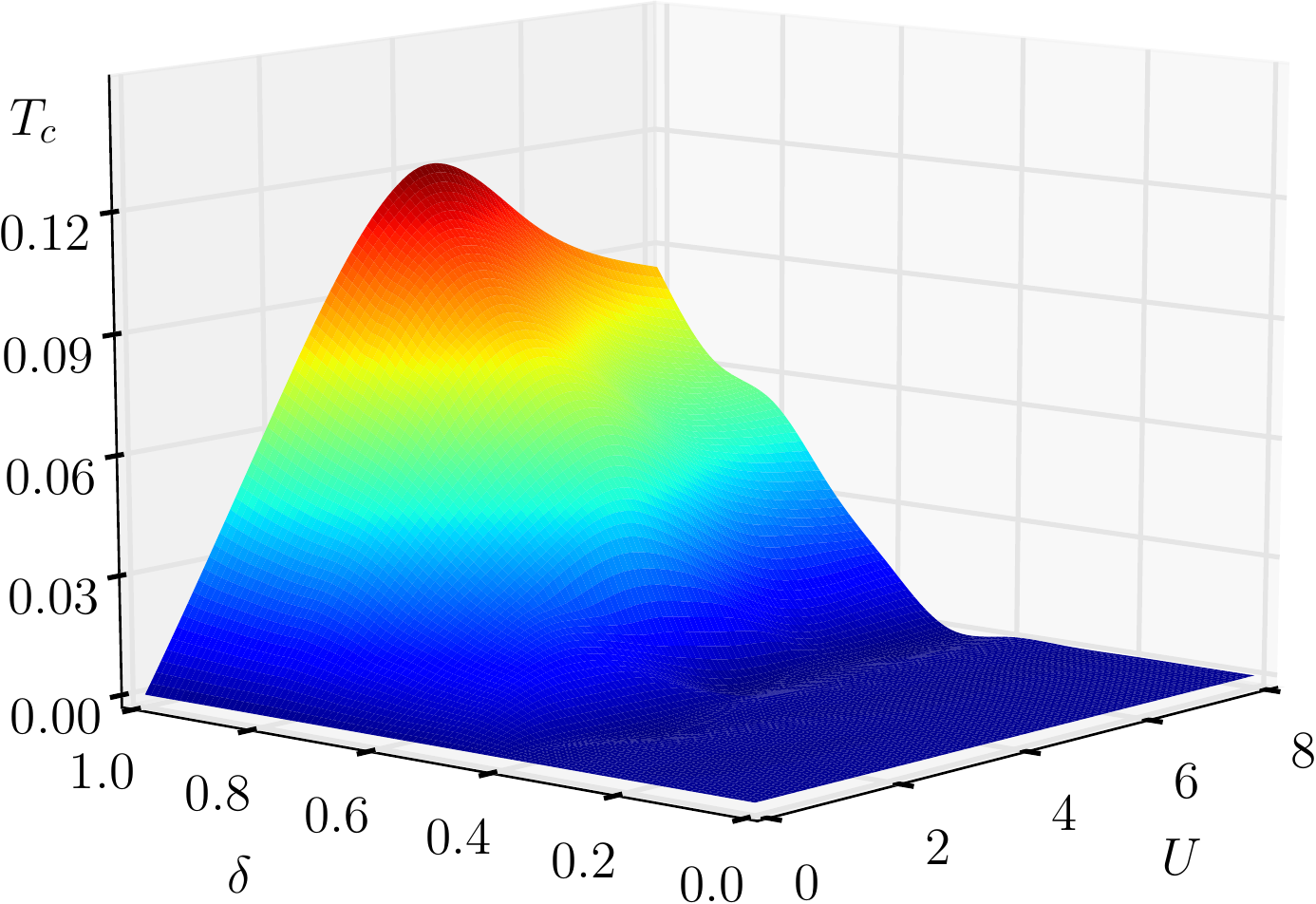}
\caption{(Color online) Phase diagram of the site-diluted attractive Hubbard model as a function of the superconducting transition temperature $T_c$, attractive interaction $U$, and
the average density of attractive centers $\delta$.}  
\end{figure}

\section{Conclusions}
In this paper, we studied random local attraction driven metal-superconductor transition   in two dimensions using  the random attractive Hubbard model.  A real space Monte Carlo method was employed after introducing 
pairing and density  fields {\it via} Hubbard-Stratanovich transformation. The method is capable of capturing the physics from weak to strong coupling regimes and gives 
a real space picture of the transition, both of which are crucial to the problem at hand. In particular, effect of intrinsic disorder on the BCS-BEC crossover has been studied.\\
~\\
The main results are as follows.  As observed in previous studies, we find that there is a critical concentration of attractive centers needed for the onset of globally phase coherent  superconductivity.  The critical concentration increases with $U$ and appears to saturate above some value. 
For small values of $U$ the transition is of percolative nature and the physics is akin to that of a BCS superconductor.
The picture that emerges is that of superconducting puddles, internally phase coherent, percolating at a critical concentration of attractive centers
 to give a globally superconducting state. However, the scenario is different at larger strengths of interaction. In this regime, the zero-temperature pairing gap continues to increase with $U$,
 local pairing tendencies persist even above the transition temperature, but a dominant role is played by strong spatial phase fluctuations, resulting in a BCS-BEC crossover. Transition temperatures are suppressed  even though there is a robust spectral gap due to local pairing and $T_c$ shows a nonmonotonic behavior as $U$ is varied. The high temperature normal state
 transforms from metal to a gapped phase as $U$ increases  and has pseudogap features in the intermediate region  due to the existence of short-range order of the amplitude of the order parameter.  Spectral functions and transport properties corroborate these findings.\\
~\\
Next, we comment on a few shortcomings of our study. First, we have allowed for numerically exact treatment of thermal fluctuations of the order parameter (and also, of the charge field)
but neglected their dynamics. Thus quantum fluctuations of the order parameter are not taken into account. This may affect the low temperature properties,
 especially near the critical concentration of impurities.  Second, while the method works very well away from half filling, 
 it is not expected to capture the charge density wave (CDW) instability at half filling that competes with superconductivity. One expects that $T_c$ will  reduce 
 to zero at larger values of $U$ and the system will transform to  a CDW phase. The present method allows for large charge fluctuations at sites as $U_i$ varies. This calls for caution. In real systems, there will be long-range Coulomb interaction, which would not allow such charge fluctuations as it costs Coulomb energy.  We wish to address this problem in future. In real two-dimensional systems there  cannot be finite temperature phase transitions, in contrast to what is seen here, since thermal fluctuations prohibit any order at nonzero temperatures. 
 However, $T_c$ shown here should be thought of as a sort of crossover scale below 
which correlation length increases rapidly. If so, even a weak coupling to a third dimension will stabilize the SC phase at nonzero temperatures.\\
~\\
The present work could be extended in many ways, some of which are currently underway.
One could include the dynamics of order parameter field in a semiclassical way after obtaining the equilibrium configurations. 
This would allow us to extract some interesting physical properties in the normal phase having preformed pairs at moderate-to-large coupling, reminiscent of the Nernst effect in cuprates,
 and its persistence in the BEC regime.   One could study the competition between disorder arising due to site dilution with those of alternate origin, for example, the presence of magnetic impurities.  The latter is expected to have a detrimental effect on the BCS state. The effect of site dilution in an insulating host is an interesting problem where an insulating gap and SC gap compete with each other \cite{pistolesi}; the application of the present method shows a further suppression of $T_c$
 as one moves towards the clean limit ($\delta \sim$ 1) and a transition to a charge-modulated insulating phase\cite{pradhan}. A further extension would be to study the role of site dilution in SC systems with order parameter symmetry different from the $s$-wave considered here as well as in imbalanced lattice fermion system\cite{FFLO} that could support breached pair and Fulde-Ferrel-Larkin-Ovchinnikov (FFLO) states.   

~\\

\section{Acknowledgements}
We acknowledge the use of high performance computing clusters at HRI.


\begin{thebibliography}{100}
\bibitem{note1} In some cases doping may induce local structural changes in the insulating host, promoting  pairing.  Impurity band superconductivity  is another possibility. The simplified model in Eq. (1) does not capture such refinements.
\bibitem{note2} In practice, we take  $U_i$ to be a negligibly small number  $ (\sim 10^{-5} )$ with $P(U) = 1-\delta$.
\end{thebibliography}
\end{document}